\documentclass{emulateapj}
\newcommand{\bjdtdb}{${\rm {BJD_{TDB}}}$}

\newcommand{\teff}{{T_{\rm eff}}}

\newcommand{\msun}{${\rm M}_\Sun$}
\newcommand{\rsun}{${\rm R}_\Sun$}

\newcommand{\mj}{${\,{\rm M}_{\rm J}}$}
\newcommand{\rj}{${\,{\rm R}_{\rm J}}$}
\newcommand{\fave}{\langle F \rangle}
\newcommand{\fluxcgs}{10$^9$ erg s$^{-1}$ cm$^{-2}$}
\newcommand{\three}{3.6$\mu$m\ }
\newcommand{\four}{4.5$\mu$m\ }

\begin{document}

\title{\emph{SPITZER} AND $Z'$ SECONDARY ECLIPSE OBSERVATIONS OF THE HIGHLY IRRADIATED TRANSITING BROWN DWARF KELT-1b}

\author{Thomas G.\ Beatty\altaffilmark{1}, Karen A. Collins\altaffilmark{2}, Jonathan Fortney\altaffilmark{3}, Heather Knutson\altaffilmark{4}, B. Scott Gaudi\altaffilmark{1}, Jacob M. Bruns\altaffilmark{5}, Adam P. Showman\altaffilmark{6}, Jason Eastman\altaffilmark{7}, Joshua Pepper\altaffilmark{8}, Robert Siverd\altaffilmark{9}, Keivan G. Stassun\altaffilmark{9,10}, and John F. Kielkopf\altaffilmark{2}
}

\altaffiltext{1}{Department of Astronomy, The Ohio State University, 140 W.\ 18th Ave., Columbus, OH 43210, USA; tbeatty@astronomy.ohio-state.edu}
\altaffiltext{2}{Department of Physics \& Astronomy, University of Louisville, Louisville, KY 40292, USA}
\altaffiltext{3}{Department of Astronomy and Astrophysics, University of California, Santa Cruz, CA 95064, USA}
\altaffiltext{4}{Division of Geological and Planetary Sciences, California Institute of Technology, Pasadena, CA 91125, USA}
\altaffiltext{5}{Department of Astrophysical \& Planetary Sciences, University of Colorado, Boulder, CO 80309, USA}
\altaffiltext{6}{Lunar and Planetary Laboratory, 1629 E. University Blvd., University of Arizona, Tucson, AZ, USA}
\altaffiltext{7}{Las Cumbres Observatory Global Telescope Network, 6740 Cortona Dr., Suite 102, Santa Barbara, CA 93117, USA}
\altaffiltext{8}{Department of Physics, Lehigh University, Bethlehem, PA, 18015, USA}
\altaffiltext{9}{Department of Physics and Astronomy, Vanderbilt University, Nashville, TN 37235, USA}
\altaffiltext{10}{Physics Department, Fisk University, 1000 17th Ave. North, Nashville, TN 37208, USA}

\shorttitle{KELT-1b Secondary Observations}
\shortauthors{Beatty et al.}

\begin{abstract}
We present secondary eclipse observations of the highly irradiated transiting brown dwarf KELT-1b. These observations represent the first constraints on the atmospheric dynamics of a highly irradiated brown dwarf, the atmospheres of irradiated giant planets at high surface gravity, and the atmospheres of brown dwarfs that are dominated by external, rather than internal, energy. Using the Spitzer Space Telescope, we measure secondary eclipse depths of $0.195\pm0.010$\% at \three and $0.200\pm0.012$\% at 4.5$\,\mu$m. We also find tentative evidence for the secondary eclipse in the $z'$ band with a depth of $0.049\pm0.023$\%. These measured eclipse depths are most consistent with an atmosphere model in which there is a strong substellar hotspot, implying that heat redistribution in the atmosphere of KELT-1b is low. While models with a more mild hotspot or even with dayside heat redistribution are only marginally disfavored, models with complete heat redistribution are strongly ruled out. The eclipse depths also prefer an atmosphere with no TiO inversion layer, although a model with TiO inversion is permitted in the dayside heat redistribution case, and we consider the possibility of a day-night TiO cold trap in this object. For the first time, we compare the IRAC colors of brown dwarfs and hot Jupiters as a function of effective temperature. Importantly, our measurements reveal that KELT-1b has a $[3.6]-[4.5]$ color of $0.07\pm0.11$, identical to that of isolated brown dwarfs of similarly high temperature. In contrast, hot Jupiters generally show redder $[3.6]-[4.5]$ colors of $\sim$0.4, with a very large range from $\sim$0 to $\sim$1. Evidently, despite being more similar to hot Jupiters than to isolated brown dwarfs in terms of external forcing of the atmosphere by stellar insolation, KELT-1b appears to have an atmosphere most like that of other brown dwarfs. This suggests that surface gravity is very important in controlling the atmospheric systems of substellar mass bodies.
\end{abstract}

\section{Introduction}

Among substellar objects, the relationship between giant planets and brown dwarfs is unclear. The generally acknowledged dividing line between these two classes of objects is based on mass. Specifically, objects above the minimum mass to burn deuterium are defined to be brown dwarfs, whereas objects less massive than this limit are defined to be planets. The deuterium burning limit is roughly $\sim$13\mj, although in detail this depends on one's definition of ``burning deuterium,'' and on the detailed composition of the object \citep{spiegel2011}.

\begin{deluxetable*}{lcc}
\tablecaption{Relevant Discovery Paper Values}
\tablewidth{0pt}
\tabletypesize{\small}
\tablehead{\colhead{~~~Parameter} & \colhead{Units} & \colhead{Value}}
\startdata
\sidehead{Stellar Parameters:}
                                      ~~~$M_{*}$\dotfill &Mass (\msun)\dotfill & $1.335\pm{0.063}$\\
                                    ~~~$R_{*}$\dotfill &Radius (\rsun)\dotfill & $1.471_{-0.035}^{+0.045}$\\
                         ~~~$\teff$\dotfill &Effective temperature (K)\dotfill & $6516\pm49$\\
\sidehead{Planetary Parameters:}
                                         ~~~$P$\dotfill &Period (days)\dotfill & $1.217514\pm0.000015$\\
                                        ~~~$M_{P}$\dotfill &Mass (\mj)\dotfill & $27.38\pm0.93$\\
                                      ~~~$R_{P}$\dotfill &Radius (\rj)\dotfill & $1.116_{-0.029}^{+0.038}$\\
                                          ~~~$e$\dotfill &Eccentricity\dotfill & $0.0099_{-0.0069}^{+0.010}$\\
               ~~~$\omega_*$\dotfill &Argument of periastron (degress)\dotfill & $61_{-79}^{+71}$\\
                                          ~~~$e\cos\omega_*$\dotfill &\dotfill & $0.0018_{-0.0059}^{+0.0092}$\\
                                          ~~~$e\sin\omega_*$\dotfill &\dotfill & $0.0041_{-0.0062}^{+0.011}$\\
                             ~~~$\log{g_{P}}$\dotfill &Surface gravity\dotfill & $4.736_{-0.025}^{+0.017}$\\
                  ~~~$T_{\rm eq}$\dotfill &Equilibrium temperature (K)\dotfill & $2432_{-27}^{+34}$\\
                          ~~~$\fave$\dotfill &Incident flux (\fluxcgs)\dotfill & $7.83_{-0.34}^{+0.45}$\\
\sidehead{Primary Transit Parameters:}
       ~~~$R_{P}/R_{*}$\dotfill &Radius of the planet in stellar radii\dotfill & $0.07806_{-0.00058}^{+0.00061}$\\
                  ~~~$a/R_*$\dotfill &Semi-major axis in stellar radii\dotfill & $3.619_{-0.087}^{+0.055}$\\
                                 ~~~$i$\dotfill &Inclination (degrees)\dotfill & $87.6_{-1.9}^{+1.4}$\\
                                      ~~~$b$\dotfill &Impact parameter\dotfill & $0.150_{-0.088}^{+0.11}$\\
                                    ~~~$\delta$\dotfill &Transit depth\dotfill & $0.006094_{-0.000090}^{+0.000096}$\\
                           ~~~$T_{FWHM}$\dotfill &FWHM duration (days)\dotfill & $0.10645\pm0.00045$\\
                     ~~~$\tau$\dotfill &Ingress/egress duration (days)\dotfill & $0.00873_{-0.00020}^{+0.00049}$\\
                            ~~~$T_{14}$\dotfill &Total duration (days)\dotfill & $0.11526_{-0.00059}^{+0.00069}$\\
\enddata
\label{tab:KELT-1b.finalfinal.median}
\end{deluxetable*}

On the one hand, distinguishing between objects below and above 13\mj\ is clearly arbitrary, particularly since after roughly a billion years deuterium burning is over and any evidence of this initial internal energy source is largely gone, i.e., an old $\sim$50\mj\ object that never fused deuterium would be difficult to distinguish from one that did \citep{spiegel2011,bodenheimer2013,molliere2012}. Therefore, giant planets and brown dwarfs can properly be thought of as a continuum of objects, with masses and surface gravities that vary accordingly. By studying how the observable properties of these objects vary as a function of mass and surface gravity for controlled samples with similar compositions and in similar environments, we can gain insight into the uncertain physics at work in these bodies. Such insights will in turn constrain the origin of these bodies. Particularly important in this regard are constraints on the atmospheric systems of brown dwarfs and giant planets, as these systems present not only the most uncertain physics, but are also the most amenable to empirical constraints.

On the other hand, giant planets and massive brown dwarfs likely have distinct origins, at least for companions to sunlike stars. This is evidenced by the existence of the brown dwarf desert, the local minimum in the mass function of relatively close-in ($\lesssim\,$10AU) companions to sunlike stars near $\sim$30 to $\sim$50\mj\ \citep[e.g.,][]{marcy2000,grether2006,sahlmann2011}. Presumably, objects below the brown dwarf desert were formed in a circumstellar disk, whereas objects above were formed in a manner more analogous to stars. However, this hypothesis is relatively untested, and even within this interpretation many questions remain. For example: what is largest-mass object that can form in a circumstellar disk? What is the smallest mass companion that can form like a star? Do these masses overlap? Does the brown dwarf desert depend on the properties of the star, or the separation from the star? Why is there apparently no brown dwarf desert for isolated objects?  By better understanding the physics of companions from the giant planet up through the brown dwarf regime, we may better understand their origins and thus provide answers to these questions.

Unfortunately, obtaining empirical constraints on giant planets and brown dwarfs in similar environments has proven difficult. The majority of our empirical constraints on brown dwarfs come from isolated brown dwarfs, brown dwarf binaries, or brown dwarfs as wide companions to stars \citep{luhman2012}. These systems are often amenable to detailed study of their atmospheres, including spectra and time series photometry. However, in the vast majority of cases, these objects do not have masses, radii, or age measurements. One exception is the eclipsing brown dwarf system 2M0535-05 \citep{stassun2006,stassun2007}, which allows for a direct measurement of the masses and radii of the two brown dwarf components. Furthermore, analogous isolated or wide-separation giant planets are considerably more difficult to study, due to their intrinsic faintness. It is only recently, and only for relatively massive and young planetary-mass objects, that the first such empirical constraints have been obtained \citep[][among others]{marois2008,lagrange2010,liu2013}.

As expected, obtaining the kinds of measurements routinely acquired for more massive brown dwarfs for these first planetary-mass objects has already provided important insight into the physics at work in these bodies. For example, consider the the strong J-band brightening that brown dwarfs undergo as they cross the L- to T-dwarf transition. This is widely interpreted as clouds clearing from the atmospheres of the L-dwarfs \citep{marley2010}, though the precise mechanism for this process is not well understood \citep{burgasser2013}. Direct imaging observations of the HR 8799 planets, which are giant planets at or past the L/T transition, do not show a similar J-band brightening \citep{faherty2013}. This indicates that the atmospheric dynamics responsible for the L/T transition is highly dependent on surface gravity \citep{bowler2010}. If we are able to understand what the differences are between atmospheric flows and forcing mechanisms in brown dwarfs and exoplanets, we may be able to understand what is precisely occurring at the L/T transition.

In contrast to brown dwarfs, the majority of our empirical constraints on giant planets comes from transiting systems. These systems provide masses, radii, and crude ages for most systems, but because of selection biases, nearly all these planets are on short periods and so likely are tidally locked and subject to very strong stellar irradiation, dramatically altering and complicating their atmospheres and atmospheric dynamics. Therefore, the empirical constraints on these systems cannot be directly interpreted in the same context as isolated brown dwarfs, hampering the ability to define the relationship between these two types of objects. The existence of the brown dwarf desert, and the resulting paucity of transiting brown dwarfs, has prevented any direct comparison of observations of brown dwarfs under similarly irradiated environments as close-in giant planets. 

KELT-1b \citep{siverd2012} provides the best opportunity to directly compare a brown dwarf to giant exoplanets under the same environmental condition of strong external irradiation. The previously discovered transiting brown dwarf companions to stars, CoRoT-3b \citep{deleuil2008}, CoRoT-15b \citep{bouchy2011}, Kepler-39b \citep{bouchy2011b}, WASP-30b \citep{anderson2011}, KOI-415b \citep{moutou2013}, KOI-205b \citep{diaz2013}, and LHS 6343C \citep{johnson2011}, all orbit relatively faint stars (with the brightest being WASP-30 at $V=11.9$ and the others being significantly fainter). On the other hand there are massive transiting giant planets like  HAT-P-2b \citep{bakos2007}, XO-3b \citep{johnskrull2008}, and WASP-18b \citep{hellier2009}, all with masses around 10 \mj, that transit stars bright enough to allow for high quality follow-up observations, but their masses place them within the planetary regime.

KELT-1b is a 27\mj\ object on a short 1.2 day orbit around a bright ($V=10.8$) F5V star. The close orbit places KELT-1b in a highly irradiated environment, with an incident stellar flux of $7.8\times10^9$ erg s$^{-1}$ cm$^{-2}$, that places it forty times above the empirical threshold for inflated planets determined by \cite{miller2011} and \cite{demory2011}. In addition, based on the expected tidal synchronization timescale of $\sim$10Myr for KELT-1b \citep{guillot1996} we expect that KELT-1b is tidally locked to its orbital period. KELT-1b thus allows us to study a brown dwarf where we know the mass and radius, in an irradiation and tidal environment similar to hot Jupiters, and around a star bright enough to allow for precision follow-up observations. To take advantage of the opportunity offered by KELT-1b, we observed several secondary eclipses of the KELT-1 system from the ground and from space during the fall of 2012. 

\section{Observations}
\subsection{Spitzer Observations}

\begin{figure}
\vskip -0.0in 
\epsscale{1.2} 
\plotone{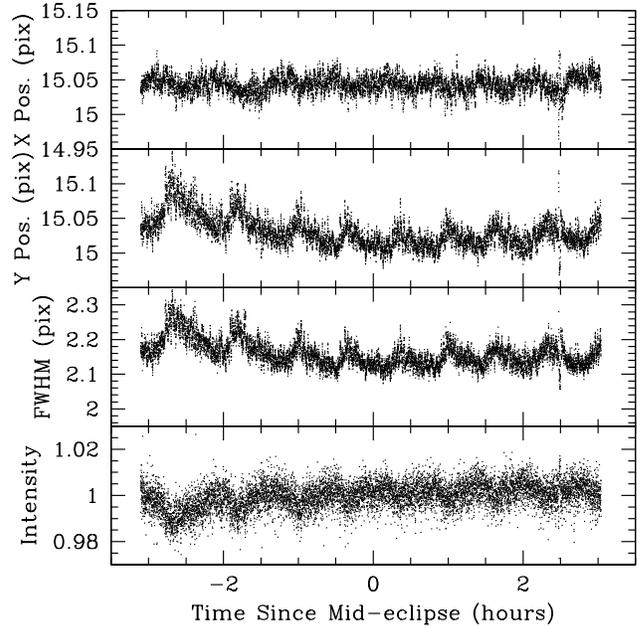}
\vskip -0.0in 
\caption{X position, Y position, FWHM, and raw aperture photometry for the \three observations. The photometry is for our chosen optimal aperture for the \three data, as described in Section 2.1.}
\end{figure}

We observed secondary eclipses of KELT-1b in the \three and \four bands using the {\sc irac} instrument on the Spitzer Space Telescope. We took the \three observations over the course of UT 2012 September 10 and 11 and the \four observations during UT 2012 September 11 and 12. Both sets of observations lasted for 6.57 hours and used the subarray mode with 2.0 second exposures. We additionally used the spacecraft's peak-up mode, with KELT-1 as the peak-up target, to enhance the pointing stability. We executed two observing sequences in each band: the first was 0.5 hours long intended to allow for the detector ramp and for the spacecraft to settle its pointing. We discarded this first observing sequence, since it was dominated by these two effects. The second sequence in each band was the main science observing sequence. This lasted for 6.07 hours and provided us with 10,944 images in each band. We used the basic calibrated data ({\sc BCD}) images from these two sequences for all our photometry.

\begin{figure}
\vskip -0.2in 
\epsscale{1.2} 
\plotone{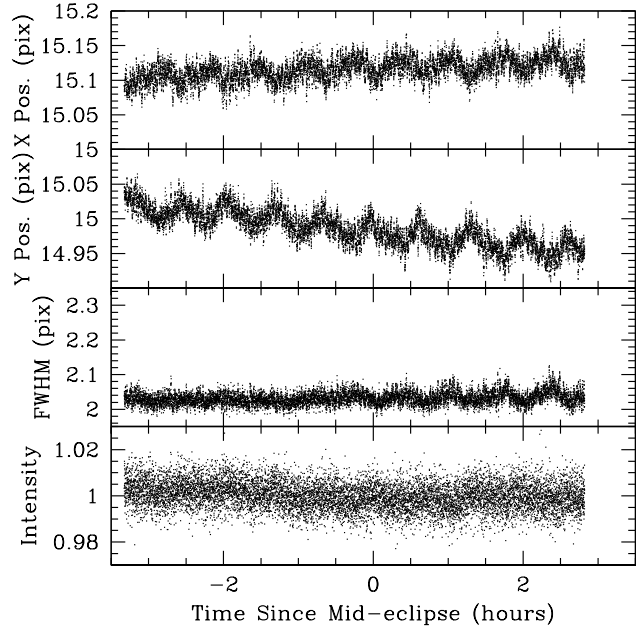}
\vskip -0.0in 
\caption{X position, Y position, FWHM, and raw aperture photometry for the \four observations. The photometry is for our chosen optimal aperture for the \four data, as described in Section 2.1.}
\end{figure}

We calculated the $\mathrm{BJD}_{\mathrm{TDB}}$ time of each image by using the {\sc fits} header entries as follows. The {\sc BCD} images come in 64 image data cubes with header keywords that record the start time ({\sc mbjd\_obs}) for each sequence containing 64 images, and the total sequence duration ({\sc aintbeg} and {\sc atimeend}). We calculated the $\mathrm{BJD}_{\mathrm{UTC}}$ time at mid-exposure for each of the 64 images in each data cube by assuming the image sequences began at {\sc mbjd\_obs} and that the 64 images in each sequence were evenly spaced between {\sc aintbeg} and {\sc atimeend}. To adhere to the timing system used in the KELT-1 discovery paper, we converted to $\mathrm{BJD}_{\mathrm{TDB}}$ by adding 64.184 seconds to the derived $\mathrm{BJD}_{\mathrm{UTC}}$ times \citep{eastman2010}.

To calculate the background in each of the 32$\times$32 pixel subarray images, we first excluded the light from the KELT-1 system by masking out a central circular area 12 pixels in radius. On each image, we then performed three rounds of 3$\sigma$ clipping on the remaining non-masked area to remove outliers. We estimate the background flux for each image by fitting a Gaussian to a histogram of the values of the remaining pixels and used the fitted value for the mean as the background flux. We subtracted this fitted mean background value from each image.  The mean background flux averaged over all the images used in the analysis was 0.05\% of KELT-1's flux at \three and 0.02\% at 4.5$\,\mu$m. 

We extracted lightcurves in each band using simple aperture photometry. To determine the position of the star in our images we fit a two-dimensional Gaussian to the stellar PSF using the entire image.  We also tried finding the star's position using flux-weighted centroiding following \cite{knutson2008}, but later found this provided inferior corrections for intrapixel sensitivity variations in our data. In doing the position determinations we used a modified set of our background-subtracted images: we replaced any hot pixels with median flux values for that pixel to prevent spurious centroid shifts. We identified pixels as ``hot'' in a particular image if their flux in the image was more than 3$\sigma$ away from the median flux for that pixel over an entire 64 image data cube. We then replaced the hot pixel with the pixel's 64 image median flux. For our aperture photometry we used the original background subtracted images without the hot pixels replaced, so as to remain as close to the raw data as possible. Instead, we used 5$\sigma$ clipping on the aperture photometry results to remove images where a hot pixel occurred within the photometric aperture.

\begin{figure}
\vskip -0.0in 
\epsscale{1.2} 
\plotone{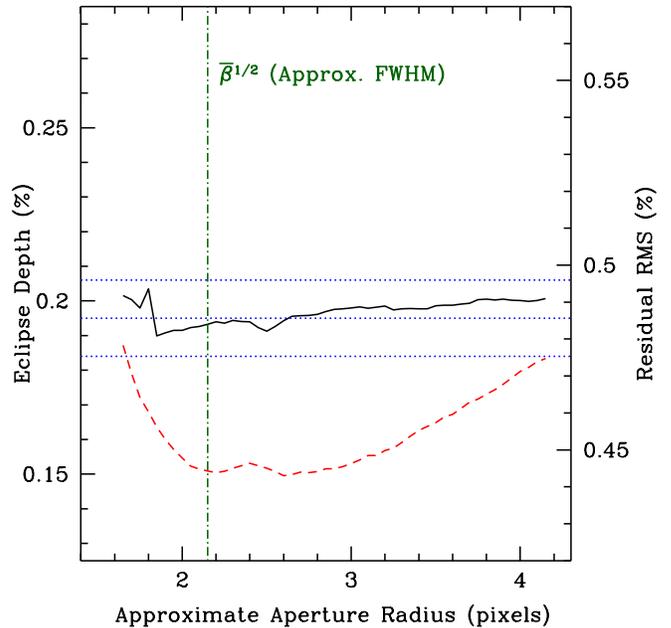}
\vskip -0.0in 
\caption{Variation of the measured eclipse depth (black, solid, and the left axis) and RMS of the residuals to the best fit (red, dashed, and the right axis) as a function of aperture size for the \three data. The blue dotted lines show our final result for the eclipse depth at \three and the 1$\sigma$ error bars. The vertical green dot-dashed line shows the median value of $\sqrt{\bar{\beta}}$, the approximate stellar FWHM.}
\end{figure}

\begin{figure}[h]
\vskip -0.2in 
\epsscale{1.2} 
\plotone{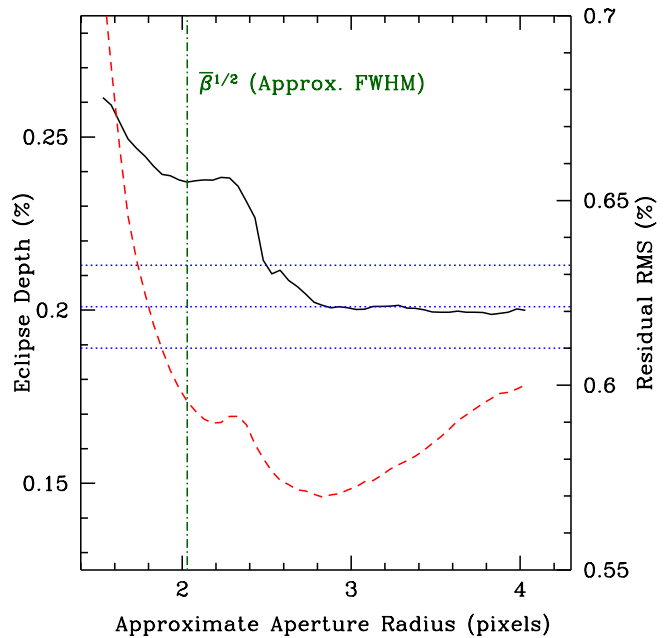}
\vskip -0.0in 
\caption{Variation of the measured eclipse depth (black, solid, and the left axis) and RMS of the residuals to the best fit (red, dashed, and the right axis) as a function of aperture size for the \four data. The blue dotted lines show our final result for the eclipse depth at \four and the 1$\sigma$ error bars. We were not able to adequately determine the cause of the bump around 2.3 pixels (see the end of Section 2.1 in the text). The vertical green dot-dashed line shows the median value of $\sqrt{\bar{\beta}}$, the approximate stellar FWHM.}
\end{figure}

To extract photometry from the images we chose to use variable apertures scaled to the FWHM of the stellar PSF for both the \three and \four data. Other secondary eclipse measurements with Spitzer usually use a variable aperture for \three data and a fixed aperture for their \four observations \citep[e.g.,][]{knutson2012,baskin2013}, but we found that using a variable aperture for both bands gave the lowest RMS scatter in the residuals to the best fit lightcurves. We estimated the width of the stellar PSF in each image using the noise pixel parameter \citep[following][]{knutson2012}, which is defined in Section 2.2.2 of the {\sc irac} instrument handbook as
\begin{equation}\label{eq:2010}
\bar{\beta} = \frac{(\Sigma_i\; I_i)^2}{\Sigma_i\; I_i^2},
\end{equation}
where $I_i$ is the intensity of the $i$th pixel. To calculate the noise pixel parameter in our data we summed over the pixels within a radius of three pixels of KELT-1's position in the image, including fractional pixels. The FWHM of the stellar PSF is then $\sqrt{\bar{\beta}}$ \citep{mighell2005}. The third panels in Figures 1 and 2 show how the FWHM calculated from the noise pixel parameter (i.e. $\sqrt{\bar{\beta}}$) varied as a function of time. At \three the median FWHM over the 9,775 images we used was 2.15 pixels, with a standard deviation of 0.04 pixels. At \four the median over the 10,307 images we used was 2.03 pixels with a standard deviation of 0.02 pixels. At \three the noise pixel parameter was nearly perfectly ($r=0.95$) correlated with the y-position of the stellar centroid. 

We set our photometric aperture size to be $\sqrt{\bar{\beta}}+C$, which is the FWHM, $\sqrt{\bar{\beta}}$, in a particular image plus some constant, $C$. We chose the optimum aperture by extracting photometric timeseries for a range of values of $C$, fitting an eclipse model to each of these lightcurves, and then choosing the value of $C$ which resulted in the lowest RMS residuals with respect to the model. We tested values $C$ from $-0.5$ pixels to $+2.0$ pixels in steps of 0.05. This roughly corresponds to apertures with radii of 1.6 to 4.1 pixels (Figures 3 and 4), but we remind the reader that $\sqrt{\bar{\beta}}$ varies with time. The approximate radii plotted in Figures 3 and 4 use the median values for $\sqrt{\bar{\beta}}$ over all images. We fit the photometry using the first \textsc{amoeba} stage of the fitting procedure described in Section 4. Figures 3 and 4 show the fitted depth (black) and residual RMS (red) as a function of aperture size for the \three and \four data.

The lowest residual RMS occurred for an aperture size of $\sqrt{\bar{\beta}}+0.5$ pixels in the \three data and $\sqrt{\bar{\beta}}+0.8$ pixels in the \four data. We therefore utilized these aperture sizes to extract the photometry that we employ in our final analysis. The standard deviation on fitted depth as a function of aperture size in the \three data was 0.003\% over the entire trial range, which is below the 0.011\% final uncertainty we find for the \three eclipse. For the \four data this was not the case, so we considered the behavior of this dataset in more detail.

At \four we found the fitted eclipse depth varied by 0.06\%, depending upon the aperture sized used to extract the photometry (Figure 4), particularly for apertures smaller than $\sqrt{\bar{\beta}}+0.7$. This is significantly above the 0.012\% final uncertainty we calculate for the \four data. There is also a ``bump'' in the RMS and fitted depth around an aperture radius of $\sqrt{\bar{\beta}}+0.25$. For reference, we refer the reader to Figure 6 of \cite{blecic2014} for an example of ``well-behaved'' \four data. We were not able to satisfactorily determine the cause of the variation, or the reason for the bump. We first tried switching the lightcurve extraction to use a non-scaled aperture size, instead of an aperture scaled to the FWHM of the stellar PSF, but these lightcurves showed a similar variability in the eclipse depth and a ``bump'' at a radius of 2.3 pixels. Next, we tested to see if the bump was caused by a bad pixel by setting individual pixels in all the images to zero one-by-one. By setting pixels (14,14), (14,15) and (14,16) to zero we were able to remove the ``bump'', but this almost doubled the variability in the fitted eclipse depth for the photometry using both non-scaled and scaled aperture sizes. A visual inspection of these three pixels' timeseries showed no obvious abnormalities. Zeroing out other pixels had no discernible effect.

We ultimately decided to use an aperture of $\sqrt{\bar{\beta}}+0.8$ pixels to extract the \four photometry from the unaltered {\sc BCD} images. In Figure 4 one can see that for apertures larger than $\sqrt{\bar{\beta}}+0.7$ ($\sim2.7$) pixels the fitted eclipse depth is nearly constant, and the lowest residual RMS occurs at $\sqrt{\bar{\beta}}+0.8$. We therefore judged that whatever the cause of the systematic changes in the depth and RMS variation for smaller aperture size is mitigated for apertures larger than $\sqrt{\bar{\beta}}+0.7$, and thus, that the photometry at our chosen aperture of $\sqrt{\bar{\beta}}+0.8$ is representative of KELT-1b's true eclipse depth at 4.5$\,\mu$m. If this is not the case, then the uncertainty on the \four depth we report is an underestimate.     

We also trimmed out some of the initial images due to the remains of the initial photometric ramp. The first 1,000 images of the \three data and the first 600 \four images displayed a clearly discernible ramp feature, so we excluded them from our analysis. 

Finally, we removed points that were more than 5$\sigma$ away from the median flux. The number of points that were clipped varied slightly depending on the exact aperture size (i.e. the value of $C$) used. In the apertures we chose to use for our final analysis, this clipping removed 169 (1.7\%) \three and 37 (0.4\%) \four images.

For our chosen photometric aperture for each data stream, our final lightcurve contained 9,775 images at \three covering -2.56 to +3.01 hours around the center of secondary eclipse. The final \four lightcurve contained 10,307 images and spans from -2.98 to +2.82 hours around the center of secondary eclipse.  

\subsection{Ground-based Observations}

Over the summer and fall of 2012 we observed seven secondary eclipses of KELT-1b in $z'$ at Moore Observatory, which is operated by the University of Louisville. We used the 0.6m RCOS telescope with an Apogee U16M 4K$\times$4K CCD, giving a $26'$$\times$$26'$ field of view and a plate scale of $0''.39$ pixel$^{-1}$. Since KELT-1 is separated from its nearest detectable neighbor in DSS2 imagery by $\sim$$18''$, we were able to defocus the telescope to allow for longer exposures without the risk of blending from the neighbor star.

We used the same observing parameters for the ground-based observations across all nights. The exposure time was 240 seconds (plus a 20 second readout time), and we slightly defocused telescope to give a toroid shaped point spread function (PSF). The target and comparison stars were placed at the same detector locations, and the guiding maintained this placement within a few pixels across all nights. Our image calibration consisted of bias subtraction, dark subtraction, flat-field division, and detector non-linearity compensation. We extracted differential aperture photometry from the calibrated images using AstroImageJ (\textsc{aij}, K. A. Collins, et al., in preparation). The comparison stars were selected from sources on the detector which had $z'$ band brightness similar to KELT-1 and which produced relatively flat light curves (after airmass detrending) when compared to the other stars in the ensemble. The final comparison ensemble included four stars near KELT-1 (TYC 2785-2151-1, TYC 2781-2231-1, LTT 17089, and TYC 2785-1743-1). We chose to allow the photometric aperture radius to vary based on an \textsc{aij} estimate of the FWHM of the toroidal PSF. After testing values in the range of 1.0 to 1.4 times the estimated FWHM, we found that a factor of 1.25 minimized the scatter in the light curves. This factor resulted in an aperture radius that varied between ~20-30 pixels across the four nights. The sky background was estimated from an annulus with inner radius 40 pixels and outer radius 80 pixels. Iterative 2$\sigma$ clipping was first performed to remove outliers and stars from the background annulus. The mean of the remaining pixels was adopted as the sky background value and subtracted from each pixel in the photometric aperture.

Three of the events, on UT 2012 September 7, 2012 October 5 and 2012 October 12, suffered from abnormally poor seeing or interruptions by clouds. We excluded these observations from consideration. The other four secondary eclipses, on UT 2012 July 30, 2012 August 16, 2012 November 18 and 2012 November 29, were high-quality, complete observations of the eclipses. The typical per point uncertainties on these nights were 0.10\% to 0.13\%. The top four panels in Figure 9 show the lightcurves from these four good nights plotted individually, after being detrended against airmass and time as described in Section 3.3.2.

\section{Lightcurve Fitting and Results}

\subsection{Eclipse Model}

We modeled the IR data as a combination of a \cite{mandel2002} eclipse lightcurve and a set of decorrelation parameters. To make the eclipse lightcurves we used the implementation of the \cite{mandel2002} lightcurves built into \textsc{exofast} \citep{eastman2013}. We modeled the eclipse by assuming KELT-1b was a uniformly bright disk, with no limb-darkening, being occulted by the much larger KELT-1. Compared to a transit lightcurve, this has the immediate effect that $R_P/R_*$ and the eclipse depth are no longer directly related.

\begin{figure}
\vskip -0.0in 
\epsscale{1.2} 
\plotone{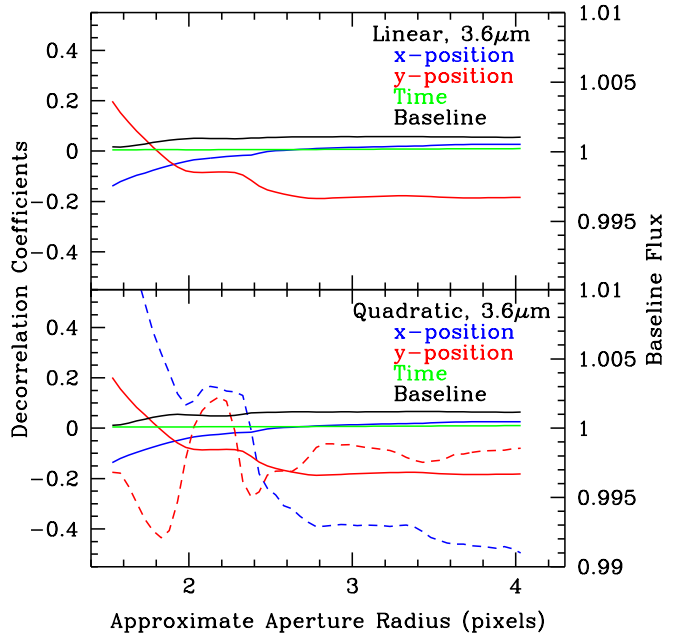}
\vskip -0.0in 
\caption{Best fit linear and quadratic decorrelation parameters for the \three data as a function of aperture radius. The black line shows the variation in the fitted baseline flux (i.e., $F_0$ in Equation \ref{eq:3110}) and is measured on the right axis, while the dashed blue and red lines show the quadratic decorrelation parameters for x- and y-position, respectively, and are measured on the left axis. The variation in the quadratic parameters indicates that they are not well constrained by our data. We use a linear decorrelation for our fits.}\vskip 0.08in 
\end{figure}

In both channels the data showed strong correlations between the flux and the $x$- and $y$-position of the star. These correlations persist regardless of the aperture size we used for the data reduction. These light curve systematics are a result of the well-known intrapixel sensitivity variation in the \three and \four detectors \citep[e.g.,][]{ballard2010}.  We fit for, and removed, these trends by including a decorrelation function that modifies the flux and contains three terms: a linear term each for the x- and y-pixel position of KELT-1, and a linear time term. We included this additional linear decorrelation against time as our initial fits using only the positional decorrelation showed a clear residual trend with time.

We considered using additional quadratic decorrelation terms in $x$- and $y$-position but found that the corresponding decorrelation coefficients varied substantially with choice of the aperture size used to extract the light curves (Figures 5 and 6). The linear terms, on the other hand, settled to specific values once the aperture size grew larger than approximately 2.6 pixels. We therefore considered the quadratic terms to be poorly unconstrained by the data. Furthermore, we found that the fitted eclipse depth was strongly correlated with the quadratic fit parameters, such that the eclipse depth was artificially suppressed when these terms were included. This is due to the relatively small amount of out-of-eclipse data, and the fact that the $x$- and $y$-position measurements are strongly correlated with time, which allowed the quadratic position terms to ``fit out'' the transit without significantly worsening the fit outside of eclipse.  

\begin{figure}
\vskip -0.0in 
\epsscale{1.2} 
\plotone{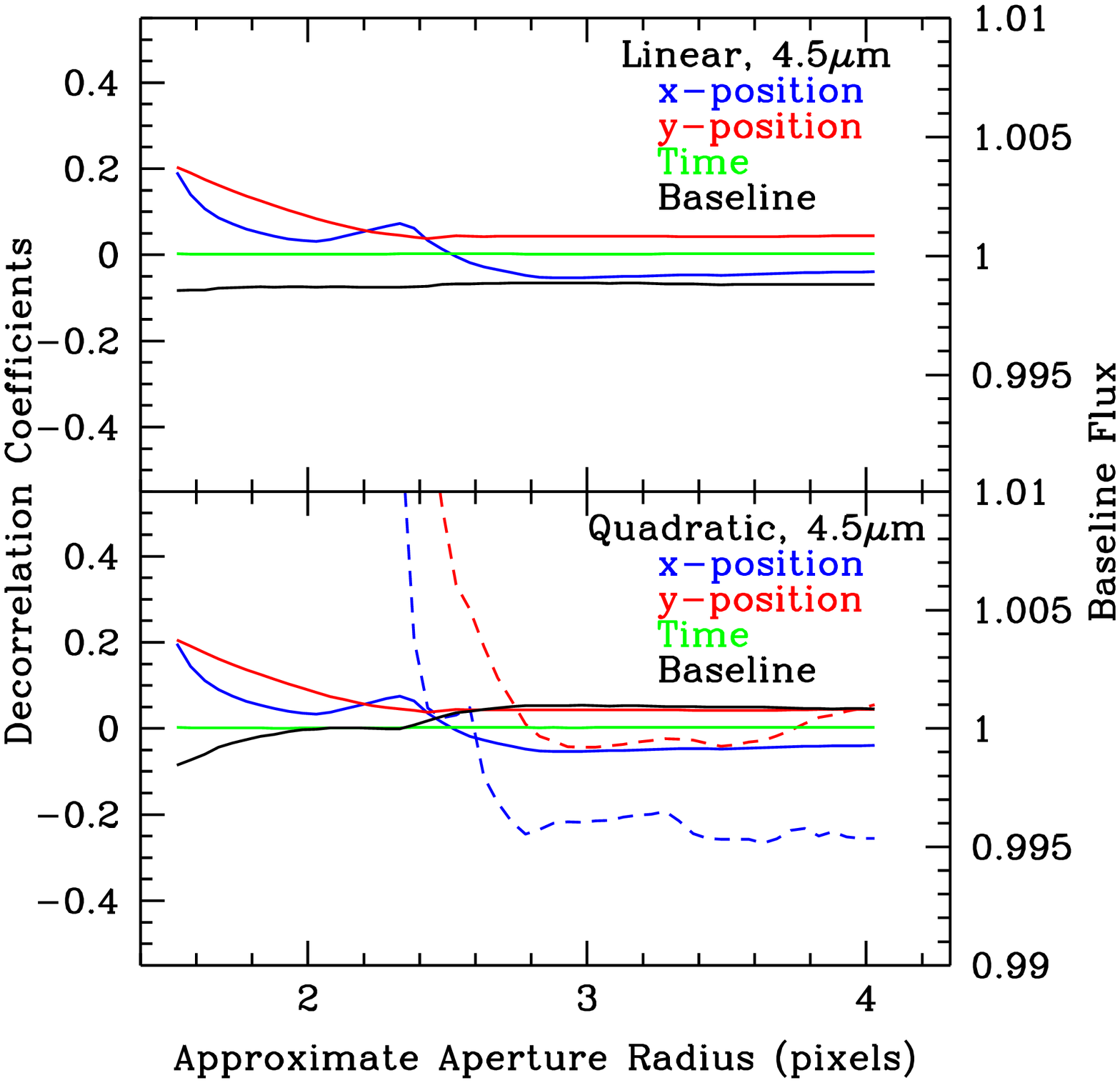}
\vskip -0.0in 
\caption{Best fit linear and quadratic decorrelation parameters for the \four data as a function of aperture radius. The black line shows the variation in the fitted baseline flux (i.e., $F_0$ in Equation \ref{eq:3110}) and is measured on the right axis, while the dashed blue and red lines show the quadratic decorrelation parameters for x- and y-position, respectively, and are measured on the left axis. The variation in the quadratic parameters indicates that they are not well constrained by our data. We use a linear decorrelation for our fits.\\}
\end{figure}

The model we fit to the data was therefore
\begin{eqnarray}\label{eq:3110}
F_m(t) &=& F_0 G(t,x,y;a_1,a_2,a_3)\\ \nonumber
&&[1-f(t;T_{C,tran},P,\sqrt{e}\cos\omega,\sqrt{e}\sin\omega,\cos i,\\ \nonumber
&&\ \ \ \ \ \ \ \ \,R_P/R_*,a/R_*,\delta)]
\end{eqnarray}
where
\begin{equation}\label{eq:3120} 
G(x,y,t;a_1,a_2,a_3) = 1 + a_1x + a_2y + a_3t
\end{equation}
is the decorrelation function which describes the variation of the unocculted total flux due to systematic effects, and $f(t)$ describes the fractional flux decrement during the eclipse as computed using the modified \cite{mandel2002} model. In addition to a time dependence, the eclipse model's exact form depends upon: the time of the previous transit ($T_{C,tran}$), the orbital period $P$, $\sqrt{e}\cos\omega$, $\sqrt{e}\sin\omega$, the cosine of the orbital inclination ($\cos i$), the radius of the planet in stellar radii ($R_P/R_*$), the semi-major axis in units of the stellar radii ($\log(a/R_*)$), the baseline flux level ($F_0$), and the eclipse depth ($\delta$).

Note that we do not explicitly fit for the time of secondary eclipse. Instead, we calculate the secondary eclipse time based on the time of the previous transit ($T_{C,tran}$), the orbital period $P$, and $\sqrt{e}\cos\omega$ and $\sqrt{e}\sin\omega$. We begin by determining the eccentricity and orientation of the orbit via $e=(\sqrt{e}\cos\omega)^2+(\sqrt{e}\sin\omega)^2$ and $\omega=\tan^{-1}(\sqrt{e}\sin\omega/\sqrt{e}\cos\omega)$. This allows us to calculate the mean anomaly of KELT-1b during transit ($M_C$) and eclipse ($M_S$). Then the eclipse time is
\begin{equation}\label{eq:3130}
T_S = T_{C,tran} + P(M_S - M_C),
\end{equation}
We explain the motivation for using this parameterization below.  

\subsection{Fitting Parameters and Their Priors}

We fit for a total of twelve parameters: the nine eclipse parameters and the three decorrelation parameters. For seven of these parameters ($T_{C,tran}$, $P$, $\sqrt{e}\cos\omega$, $\sqrt{e}\sin\omega$, $\cos i$, $R_P/R_*$, and $a/R_*$), we had a prior expectation for their values from the KELT-1b discovery paper. We did not have any prior expectations for the five remaining parameters ($F_0$, $\delta$, and the decorrelation terms). To incorporate our priors into the fitting process, we added a term for each parameter to the $\chi^2$ function of the form
\begin{equation}\label{eq:3210}
\Delta\chi^2_a = \left(\frac{a_i-a_0}{\sigma_a}\right)^2.
\end{equation}
Here $a_i$ is the trial value of an individual parameter $a$, $a_0$ is the prior value, and $\sigma_a$ is the $1\sigma$ uncertainty in that prior value. Note that this does not consider any possible covariance between the parameters. We used central values and $1\sigma$ uncertainties for $T_{C,tran}$, $P$, $\sqrt{e}\cos\omega$, $\sqrt{e}\sin\omega$, $\cos i$, $R_P/R_*$, and $a/R_*$ from the discovery paper fit that was based on a free eccentricity (see Tables 4 and 5 in \citealt{siverd2012}), which we list for convenience in Table 1. To calculate $T_{C,tran}$, the time of the previous transit, we assumed no variation in the transit times, such that $T_{C,tran}=T_C + nP$ and $\sigma^2_{T_{C,tran}}=\sigma^2_{T_C}+n^2\sigma^2_P$. We also derive values and uncertainties for $\sqrt{e}\cos\omega$ and $\sqrt{e}\sin\omega$ by using the values and uncertainties for $e$, $e\cos\omega$ and $e\sin\omega$ in Tables 4 and 5 of \cite{siverd2012} and dividing the two latter quantities by $\sqrt{e}$. We chose to use $\sqrt{e}\cos\omega$ and $\sqrt{e}\sin\omega$ as the prior parameters in our MCMC fits, rather than $e\cos\omega$ and $e\sin\omega$, as the former parameterization results in a uniform prior for $e$, while the latter leads to a prior that is proportional to $e$. 

We chose to use $T_{C,tran}$, $P$, $\sqrt{e}\cos\omega$ and $\sqrt{e}\sin\omega$ to calculate the secondary eclipse time, instead of using the predicted time of secondary eclipse, $T_S$, from Table 5 of \cite{siverd2012}, so that we could properly allow for the possibility of a non-zero eccentricity and calculate appropriate uncertainties. If we were to only fit for the time of secondary eclipse $T_S$, and calculate the orbital eccentricity based on the eclipse time and duration, then we would incorrectly be assuming a circular orbit in our modeling of the eclipse orbital geometry and lightcurve. This would mean that our priors on $\cos i$ and $a/R_*$, the two terms that would then completely set the eclipse duration, would potentially dominate our measurement of the eclipse duration and thence $\sqrt{e}\sin\omega$. On the other hand, if we allowed for eccentric eclipse geometries in the lightcurve modeling by instead fitting for $T_S$, $\sqrt{e}\cos\omega$ and $\sqrt{e}\sin\omega$ we would be double-counting the uncertainties in $\sqrt{e}\cos\omega$ and $\sqrt{e}\sin\omega$; the uncertainty on $T_S$ in Table 5 of \cite{siverd2012} already includes the uncertainties in $T_C$, $\sqrt{e}\cos\omega$ and $\sqrt{e}\sin\omega$. Using $T_{C,tran}$, $\sqrt{e}\cos\omega$ and $\sqrt{e}\sin\omega$ therefore allows us to correctly model possible eclipse geometries and compute proper uncertainties. Note again, though, that this makes the assumption that KELT-1b has a fixed orbital period and does not display transit or eclipse timing variations. We do account for the 25 second R\o mer delay the eclipse has relative to the transit ephemeris. 

As we have no prior expectation for the values of $F_0$, $\delta$, and the decorrelation terms ($a_1, a_2$ and $a_3$), we do not include a $\chi^2$ penalty for these terms, and implicitly assume a uniform prior for all five parameters. Of the twelve fitting parameters, these five are therefore the only ones set entirely by our secondary eclipse observations. 

\begin{figure}
\vskip -0.0in 
\epsscale{1.2} 
\plotone{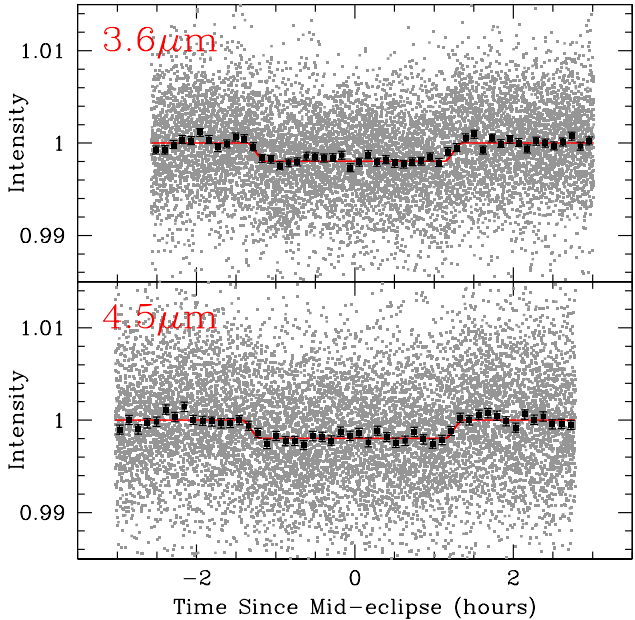}
\vskip -0.0in 
\caption{Our final, detrended, \three and \four lightcurves. The black overplotted points with error bars are the detrended data median binned into 50 points, while the red solid lines show our final best fit models to the eclipses.}
\end{figure}

Among the remaining seven terms, all of which had well defined priors, $\sqrt{e}\cos\omega$ and $\sqrt{e}\sin\omega$ are the only ones for which the data provide a tighter constraint than provided by their prior distributions. $T_{C,tran}$ and $P$ are completely determined by their priors, as the data from an individual eclipse provides no constraint on the nearest transit time or the orbital period. Similarly, our posterior constraints on  $\cos i$, $R_P/R_*$, and $\log(a/R_*)$ are dominated by our priors. This is a result of the fact that, for an eclipse lightcurve, the only constraint on $R_P/R_*$ comes from the ingress and egress durations, in contrast to a transit lightcurve, where $R_P/R_*$ is heavily determined by the depth of the transit. This means that the two major timing measurements an eclipse lightcurve provides (the FWHM and ingress/egress durations) now must be used to constrain three different parameters. As a consequence, unique values for $\cos i$, $R_P/R_*$, and $\log(a/R_*)$ are poorly constrained by the secondary eclipse data alone, and their values and uncertainties remain nearly identical to those from the discovery paper. In principle this degeneracy is resolved by the exact shape of the ingress and egress portions of the lightcurve, but this is below the precision of our data.

\subsection{Fitting Process and Results}

\subsubsection{Spitzer}

We fit our data using the \textsc{amoeba} and MCMC routines packaged with \textsc{exofast}. We chose to fit the \three and \four data separately. Our fitting process began by using \textsc{amoeba} to find initial $\chi^2$ minima to use as estimates of the best fits. This is the same procedure we used previously, to fit the data for our determination of the optimum photometric aperture size, and allowed us to quickly find a likely minimum in the $\chi^2$ surface.

We then used MCMC to explore the parameter space around the $\chi^2$ minimum found by \textsc{amoeba} to determine uncertainties. \textsc{exofast} uses the Differential Evolution Markov Chain implementation of the MCMC algorithm, which runs multiple, simultaneous, chains to determine the correct step size and direction. In addition to verifying that we had found the global $\chi^2$ minimum, the MCMC analysis also provided us with appropriate uncertainties for all of the system and eclipse parameters.

The final system parameters and errors determined by the MCMC analyses are in Tables 2 and 3. Figure 7 shows our best fits to the detrended data.

\begin{figure}[b]
\vskip -0.0in 
\epsscale{1.2} 
\plotone{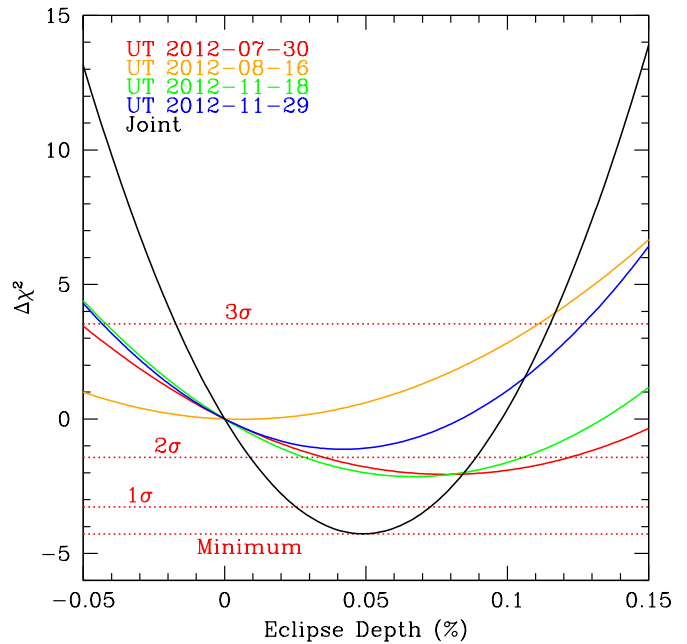}
\vskip -0.0in 
\caption{Constraints on the eclipse depth in $z'$ from our ground-based observations. Though we observed seven eclipses, only four of the nights provided high-quality, complete observations. The black line shows the joint constraint on the eclipse depth from all four nights. This assumes that the observational errors are uncorrelated from night to night.}
\end{figure}

We next conducted a prayer bead analysis on our data to assess the effects of correlated noise in the data. The prayer bead analysis we conducted followed the general description of \cite{moutou2004} and \cite{gillon2007}. In each band we took the residuals to our best fit model, shifted the residuals by one, added the incremented residuals back onto the original best fit model and then refit using \textsc{amoeba}. We shifted through the residuals to the entire lightcurve this way, such that the $i$th model point had the $i$th+$n$th residual added to it, with $n$ going from one to the number of points in the lightcurve. The remainder at the temporal end of the lightcurve was looped around and added to the beginning. 

The goal of the prayer bead analysis is to appropriately account for the presence of correlated noise in the data. This is done by examining the variation in the fitted parameters as a function of the shifting residuals. For both the \three and \four observations the standard deviations in the lightcurve and decorrelation parameters output by the prayer bead chains were within 10\% of the 1$\sigma$ error bars from the MCMC analysis. We therefore chose to use the MCMC errors as our final uncertainties for the Spitzer observations.

Another possible source of systematic uncertainty in our observations is the possible stellar companion to KELT-1 discovered by \cite{siverd2012}. The companion is located 558 mas to the southeast of KELT-1, and has $\Delta H=5.90\pm0.10$ and $\Delta K'=5.59\pm0.12$. This separation places the companion 0.47 pixels away from KELT-1 in our Spitzer observations and 1.4 pixels away in our ground-based observations. In both cases it is unresolved.

Assuming that the companion is associated with KELT-1, its luminosity difference and its $H-K'$ colors imply that it is a mid M-dwarf. If we extrapolate from $H$ and $K'$ to the Spitzer bandpasses using a Kurucz-based 6500K, $\log(g)=4.25$, spectrum for KELT-1 and a 3200K blackbody spectrum for the companion, then we find that the companion contributes less than 1\% of the total system light at \three and 4.5$\,\mu$m. In both bands this is substantially below the fractional uncertainty we calculate for the measured eclipse depths (about 10\% in both cases). This is also faint enough that the companion does not affect our measurements of KELT-1's pixel position in the images. We therefore chose to ignore its contribution to our observations.

\begin{figure}
\vskip -0.0in 
\epsscale{1.2} 
\plotone{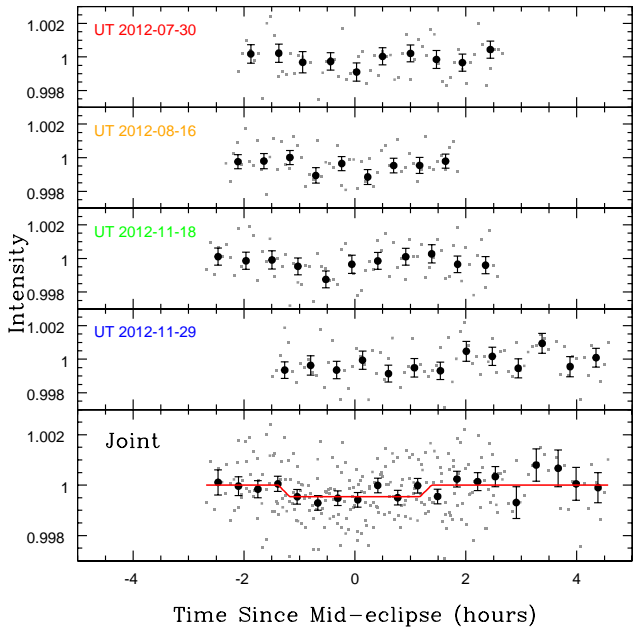}
\vskip -0.0in 
\caption{All four of the lightcurves used to calculate the constraint on the eclipse depth in $z'$. The bottom panel shows all four phased and overplotted. Each lightcurve has also been linearly detrended against airmass and time. The black points are binned versions of the individual and combined lightcurves, while the red line in the bottom panel is the best-fit eclipse model. We marginally detect the eclipse in $z'$ with a depth of $0.049\pm0.023$\%.}
\end{figure}

\subsubsection{Ground-based}

We analyzed each of the four nights individually. We fit only for the depth of a possible eclipse using a trapezoidal eclipse model that had the eclipse time, total duration and ingress/egress duration fixed. We computed the expected eclipse times for each night by extrapolating from our measured \three eclipse time and assuming a fixed period of 1.217514 days. The choice of the \three eclipse time is arbitrary; we repeated our entire analysis of the $z'$ using the \four eclipse time and found no difference in our results. The total duration and ingress/egress duration we set to the average of our \three and \four results. In addition to the eclipse model, we also included linear decorrelation terms for airmass and time. We scaled the errors on each night so that a baseline zero depth fit had a reduced $\chi^2$ of one. In all cases the scaling factor was within ten percent of unity. We used the baseline fit to calculate the $\Delta\chi^2$ values for the non-zero depth fits. 

Combining the data from all four good nights, we find suggestive evidence for an eclipse depth of $0.049\pm0.023$\% in $z'$. Figure 8 shows the $\Delta\chi^2$ as a function of eclipse depth for these four nights. The black line in Figure 8 is the $\Delta\chi^2$ of all the nights added together, and is valid under the assumption that our uncertainties are uncorrelated night to night. We have adopted this joint constraint as our final determination of the $z'$ eclipse depth of KELT-1b. Figure 9 shows the four complete $z'$ lightcurves individually, and combined, phased, and overplotted with our best fit eclipse model.

If the detection in $z'$ is real, then it is the result of thermal emission from KELT-1b, and not reflected light. In the case of an extreme Bond albedo of one, and assuming KELT-1b reflects as a Lambert sphere, then the eclipse depth due to reflected light alone would be $\sim0.03$\%. A more realistic Bond albedo of 0.1 would reduce this depth by a factor of ten, and place it far below our precision in $z'$. 

\section{Results}

We strongly detect the secondary eclipses of KELT-1b at both \three and 4.5$\,\mu$m, and weakly detect the eclipse in $z'$ (Figures 7, 9, and 10). We measure eclipse depths of $\delta_z=0.049\pm0.023$\% in $z'$, $\delta_{3.6}=0.195\pm0.010$\% at \three and $\delta_{4.5}=0.200\pm0.012$\% at 4.5$\,\mu$m. These depths correspond to brightness temperatures of $3300\,$K, $3150\,$K and $3000\,$K for the $z'$, \three and \four eclipses, respectively. 

We derive the median and 68\% confidence intervals for $e\cos\omega$, $e\sin\omega$, and $e$ from the final MCMC chain in both of the Spitzer bands. At both \three and \four we infer values of $e\cos\omega$ and $e\sin\omega$ that are consistent with zero, which signifies KELT-1b's orbit is consistent with circular. While we formally calculate a value of $e$ that is greater than zero at $>$1$\sigma$ in both bands, this is a result of the well-known Lucy-Sweeney bias \citep{lucy1971}. Our measurement of circular orbit lends credence to the strong circumstantial evidence that the orbit of KELT-1b has been tidally circularized and that the star KELT-1 has tidally synchronized to the orbital period (see Section 6.2 of the discovery paper). 

\begin{deluxetable*}{lcc}
\tablecaption{Median Values and 68\% Confidence Intervals for the \three Eclipse}
\tablehead{\colhead{~~~Parameter} & \colhead{Units} & \colhead{Value}}
\startdata
\sidehead{Measured Parameters:}
                                 $F_0$\dotfill &Baseline flux\dotfill & $1.001042\pm0.000068$\\
                 $a_1$\dotfill &X-position linear coefficient\dotfill & $0.0015\pm0.0037$\\
                 $a_2$\dotfill &Y-position linear coefficient\dotfill & $-0.1629\pm0.0028$\\
                       $a_3$\dotfill &Time linear coefficient\dotfill & $0.00623\pm0.00080$\\
             $T_C$\dotfill &Time of nearest transit (\bjdtdb)\dotfill & $2456180.7969\pm0.0022$\\
                 $\log(P)$\dotfill &Log orbital period (days)\dotfill & $0.0854741\pm0.0000053$\\
                            $\sqrt{e}\cos{\omega}$\dotfill & \dotfill & $-0.025_{-0.030}^{+0.032}$\\
                            $\sqrt{e}\sin{\omega}$\dotfill & \dotfill & $0.001_{-0.085}^{+0.072}$\\
                     $\cos{i}$\dotfill &Cosine of inclination\dotfill & $0.059_{-0.023}^{+0.020}$\\
     $R_{P}/R_{*}$\dotfill &Radius of planet in stellar radii\dotfill & $0.07807\pm0.00058$\\
$\log(a/R_{*})$\dotfill &Log semi-major axis in stellar radii\dotfill & $0.5671_{-0.0062}^{+0.0058}$\\
                              $\delta$\dotfill &Eclipse depth\dotfill & $0.00195\pm0.00010$\\
\sidehead{Derived Parameters:}
                           $P$\dotfill &Orbital period (days)\dotfill & $1.217514\pm0.000015$\\
                     $T_S$\dotfill &Time of eclipse (\bjdtdb)\dotfill & $2456181.40403_{-0.00096}^{+0.00075}$\\
          $a/R_{*}$\dotfill &Semi-major axis in stellar radii\dotfill & $3.691\pm0.051$\\
                           $i$\dotfill &Inclination (degrees)\dotfill & $86.6\pm1.2$\\
                                $b$\dotfill &Impact Parameter\dotfill & $0.220_{-0.084}^{+0.070}$\\
                     $T_{FWHM}$\dotfill &FWHM duration (days)\dotfill & $0.1037\pm0.0020$\\
               $\tau$\dotfill &Ingress/egress duration (days)\dotfill & $0.00874_{-0.00024}^{+0.00029}$\\
                      $T_{14}$\dotfill &Total duration (days)\dotfill & $0.1125\pm0.0020$\\
                                   $e\cos{\omega}$\dotfill & \dotfill & $-0.0018_{-0.0031}^{+0.0020}$\\
                                   $e\sin{\omega}$\dotfill & \dotfill & $0.0000_{-0.0077}^{+0.0059}$\\
                            $e$\dotfill &Orbital Eccentricity\dotfill & $0.0050_{-0.0036}^{+0.0081}$\\
           $\omega$\dotfill &Argument of periastron (degrees)\dotfill & $10_{-140}^{+120}$
\enddata
\label{tab:Ch. 1 KELT-1 Sec. 0.50 (variable)}
\end{deluxetable*}

\begin{deluxetable*}{lcc}
\tablecaption{Median Values and 68\% Confidence Intervals for the \four Eclipse}
\tablehead{\colhead{~~~Parameter} & \colhead{Units} & \colhead{Value}}
\startdata
\sidehead{Measured Parameters:}
                                 $F_0$\dotfill &Baseline flux\dotfill & $1.000885_{-0.000079}^{+0.000076}$\\
                 $a_1$\dotfill &X-position linear coefficient\dotfill & $-0.0524\pm0.0044$\\
                 $a_2$\dotfill &Y-position linear coefficient\dotfill & $0.0434\pm0.0039$\\
                       $a_3$\dotfill &Time linear coefficient\dotfill & $0.0020\pm0.0013$\\
             $T_C$\dotfill &Time of nearest transit (\bjdtdb)\dotfill & $2456182.0142\pm0.0023$\\
                 $\log(P)$\dotfill &Log orbital period (days)\dotfill & $0.0854739\pm0.0000053$\\
                            $\sqrt{e}\cos{\omega}$\dotfill & \dotfill & $-0.038_{-0.029}^{+0.034}$\\
                            $\sqrt{e}\sin{\omega}$\dotfill & \dotfill & $0.027_{-0.082}^{+0.074}$\\
                     $\cos{i}$\dotfill &Cosine of inclination\dotfill & $0.044\pm0.023$\\
     $R_{P}/R_{*}$\dotfill &Radius of planet in stellar radii\dotfill & $0.07806_{-0.00060}^{+0.00058}$\\
$\log(a/R_{*})$\dotfill &Log semi-major axis in stellar radii\dotfill & $0.5600\pm0.0065$\\
                              $\delta$\dotfill &Eclipse depth\dotfill & $0.00200\pm0.00012$\\
\sidehead{Derived Parameters:}
                           $P$\dotfill &Orbital period (days)\dotfill & $1.217514\pm0.000015$\\
                     $T_S$\dotfill &Time of eclipse (\bjdtdb)\dotfill & $2456182.62027_{-0.0015}^{+0.00099}$\\
          $a/R_{*}$\dotfill &Semi-major axis in stellar radii\dotfill & $3.631\pm0.054$\\
                           $i$\dotfill &Inclination (degrees)\dotfill & $87.5\pm1.3$\\
                                $b$\dotfill &Impact Parameter\dotfill & $0.160_{-0.082}^{+0.077}$\\
                     $T_{FWHM}$\dotfill &FWHM duration (days)\dotfill & $0.1063\pm0.0022$\\
               $\tau$\dotfill &Ingress/egress duration (days)\dotfill & $0.00877_{-0.00022}^{+0.00026}$\\
                      $T_{14}$\dotfill &Total duration (days)\dotfill & $0.1151\pm0.0023$\\
                                   $e\cos{\omega}$\dotfill & \dotfill & $-0.0032_{-0.0033}^{+0.0030}$\\
                                   $e\sin{\omega}$\dotfill & \dotfill & $0.0014_{-0.0055}^{+0.0096}$\\
                            $e$\dotfill &Orbital Eccentricity\dotfill & $0.0065_{-0.0045}^{+0.0081}$\\
           $\omega$\dotfill &Argument of periastron (degrees)\dotfill & $97_{-230}^{+43}$
\enddata
\label{tab:Ch. 2 KELT-1 Sec. 0.80 (variable)}
\end{deluxetable*}

\section{Discussion}

Overall, KELT-1b is a unique object: it is a relatively high mass and high surface gravity object that orbits only 3.6 stellar radii away from its host star. Among high mass sub-stellar objects, this places KELT-1b squarely in a radiation environment that until now has been populated solely by hot Jupiters. KELT-1b can therefore be interpreted in the context of a hot Jupiter with very high surface gravity, or in the context of a brown dwarf subject to strong external irradiation.  We consider both perspectives.

\subsection{From a Giant Planet Perspective}

If considered as a planet, KELT-1b stands out due to its extremely high surface gravity ($\log(g)=4.74$), which is thirty times higher than for a typical hot Jupiter. KELT-1b therefore allows us test theories of hot Jupiter atmospheres at a very high surface gravity. Of particular interest are the amount of heat redistribution and the presence of a stratospheric temperature inversion within the atmosphere of KELT-1b. \cite{perez2013} have noted that planets hotter than $\sim$2000$\,$K are observed to have extremely low amounts of heat redistribution from their day- to their nightsides, presumably because the shorter radiative timescales in hotter atmospheres cause these planets to reradiate the incident stellar flux, rather than advecting it through winds to the nightside. In KELT-1b, due to its high surface gravity, the theoretical radiative timescale is relatively longer, and the theoretical advection timescale relatively shorter, than in most hot Jupiters.

Similarly, consider the presence of a stratospheric temperature inversion in the atmosphere of KELT-1b. Temperature inversions have been observed in several hot Jupiters, predominantly among those with equilibrium temperatures higher than $2000\,$K \citep[e.g.,][]{cowan2011}. \cite{hubeny2003} and \cite{fortney2008} have proposed that gas-phase TiO in an atmosphere causes temperature inversions, since it is a strong optical absorber and condenses between $1900\,$K and $2000\,$K, depending on the pressure. However, the ultimate cause of inversions, and their precise regulatory mechanisms, have not been definitively agreed upon.

Our observations show that KELT-1b does not have a high heat redistribution efficiency. The TiO, complete redistribution atmosphere model is strongly excluded by our observations, and the highest allowable redistribution efficiency, presuming the presence of TiO, would be for day-side redistribution. Following the notation of \cite{seager2010}, it is probable that $f'>1/2$, and possible that $f'\sim2/3$ (instantaneous re-emission of the incident stellar radiation). The large difference between the equilibrium temperature of KELT-1b ($2400\,$K) and the brightness temperatures we measure at all three wavelengths ($\sim$$3100\,$K) is also indicative of an extremely low amount of heat redistribution occurring in the atmosphere. This agrees with the trend noted by \cite{perez2013} that hotter planets have lower heat redistribution inefficiencies.

That being said, and as noted previously, the increased photospheric pressure in KELT-1b may complicate this interpretation. All other things being equal, the pressure level for the $\tau=1$ surface in an atmosphere is proportional to surface gravity. Thus KELT-1b should have a photosphere at a pressure $\sim$30 times deeper than on a typical hot Jupiter. Since radiative time constants tend to increase greatly with pressure \citep{iro2005,showman2008}, one would expect the radiative time constant at the photosphere would be larger for KELT-1b than for an otherwise identical low-mass hot Jupiter. Similarly, the advection timescale scales inversely with surface gravity and scale height \citep{perez2013}, which would make advection relatively quick in KELT-1b's atmosphere. These changes in the radiative and advective timescales would allow more time for faster advection within KELT-1b's atmosphere, thereby lessening the day-night temperature difference and causing a large hotspot offset from the substellar point, for a given set of irradiation conditions. For this reason it will be interesting to compare KELT-1b to lower-mass hot Jupiters that have similar irradiation levels. Specifically, orbital phase curve observations of KELT-1b would allow one to directly measure the day-night temperature difference and test the effect of KELT-1b's greater photospheric pressure. 

\begin{figure}
\vskip -0.0in 
\epsscale{1.2} 
\plotone{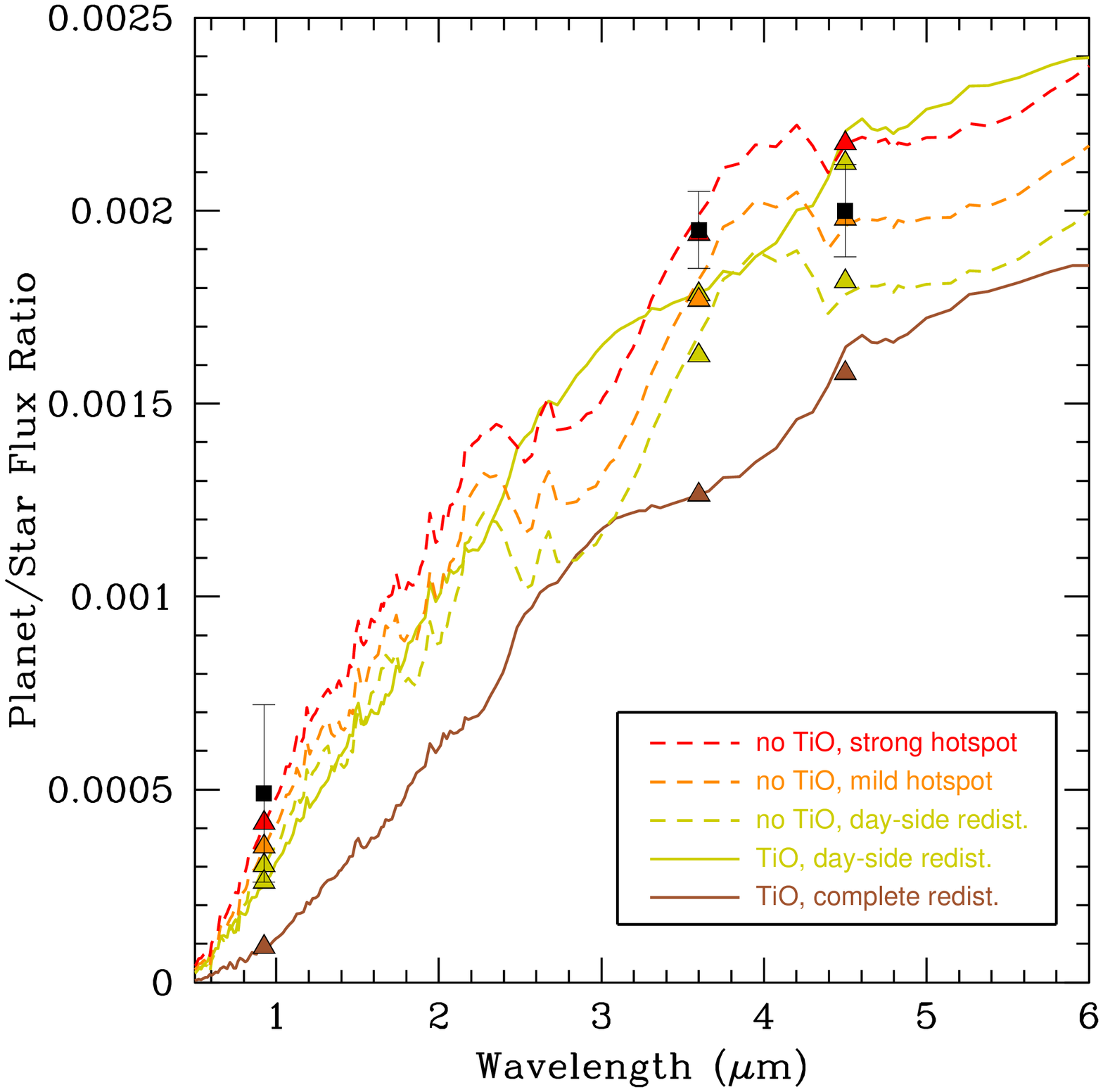}
\vskip -0.0in 
\caption{Our measured planet-to-star flux ratios at \three, \four and in $z'$. The atmosphere models are based on \cite{fortney2008}, and are divided according to the presence of gaseous TiO and the amount of heat redistribution from the day to night side. The `TiO' models have stratospheric temperature inversions, while the `no TiO' models do not. The `hotspot' models are scenarios wherein the heat from the stellar insolation is redistributed over only a portion of the planetary day side. In the `strong hotpot' this redistribution area is smaller than in the `mild hotspot' model. The colored triangles show the predicted flux ratios from each of the models in the three bandpasses..}\vskip 0.1in
\end{figure}

As a side note, our determination that $f'>1/2$ would appear to be roughly inconsistent with the ground-based secondary eclipse observations undertaken for the KELT-1b discovery paper. Figure 14 of \cite{siverd2012} implies that $f'>1/2$ would have been detectable by at least $2\sigma$ in those data. However, when we examined this issue we discovered that the analysis in \cite{siverd2012} incorrectly dealt with the observations taken using the Faulkes Telescope North (FTN). The FTN data were reported as differential magnitudes, but we treated them as differential fluxes, inverting the sense of the changes in intensity. If we recalculate Figure 14 of \cite{siverd2012} correctly, then atmospheres with $f'>1/2$ would at best be detectable at $0.8\sigma$.

Our observations are not sufficient to conclusively determine whether a TiO inversion exists in the atmosphere of KELT-1b, but we can nonetheless provide some useful constraints. Figure 10 shows our measured eclipse depths on top of atmosphere models from \cite{fortney2008}. The models without TiO (dashed lines) are for an atmosphere without an inversion, while the TiO models (solid lines) have an inversion. The best fit to the data is the no-TiO, strong hotspot model, with $\chi^2=2.23$ for three degrees of freedom. However, the no-TiO, mild hotspot model has a $\Delta\chi^2$ relative to the best model of only 1.44, while the TiO, day-side redistribution model has a $\Delta\chi^2=2.60$, and thus these models are also consistent with the data. On the other hand, the no-TiO, day-side redistribution model is marginally excluded with $\Delta\chi^2=11.31$, and while the TiO, complete redistribution model is strongly excluded with $\Delta\chi^2=60.07$. Since the $\sim$3100K brightness temperature that we measure for the day side is much hotter than the $2000\,$K condensation temperature of TiO at KELT-1b's photospheric pressure, the lack of a strong TiO signal raises the possibility that a day-night TiO cold trap exists in KELT-1b's atmosphere.

A day-night cold trap occurs for TiO when a planetary day side is hot enough to allow for gaseous TiO, but the night side is below the condensation temperature. This allows for TiO to condense and settle out of the atmosphere on the planetary nightside, removing it from the upper atmosphere. This is distinct from the cold traps predicted by 1D atmosphere models, which exist between a hot upper atmosphere and a hot lower convective layer as a pressure band cold enough to allow for gaseous TiO to condense \citep{hubeny2003,fortney2006}. Day-night cold traps were suggested as an important mechanism in hot Jupiter atmospheres by \cite{showman2009} and more thoroughly modeled by \cite{parmentier2013}, who specifically examined the role of a cold trap in HD 209458b. \cite{parmentier2013} found that TiO could settle out of the nightside atmosphere rapidly enough to prevent an inversion if the condensate grain size was larger than a few microns, which they found unlikely due to the relative scarcity of TiO. \cite{parmentier2013} did allow that if the TiO combined and condensed with a more abundant gas, such as SiO, then sufficiently large grains were much easier to form. In the case of KELT-1b, its high surface gravity may aid the efficiency of a cold-trap by increasing the particle settling velocity, and hence the settling efficiency. The dynamics of a day-night TiO cold trap may be very different on KELT-1b than on a lower surface gravity hot Jupiter.

However, as an example of the complexity of the issue, consider the presence of temperature inversions in the two other planets with surface gravities higher than $\log(g)=4.0$ that have been observed with Spitzer: WASP-18b and HAT-P-2b. WASP-18b does not show strong evidence for a temperature inversion \citep{nymeyer2011}\footnote{\cite{nymeyer2011} concludes that WASP-18b probably does have an inversion, but the authors note that their model for an inverted atmosphere is only moderately better than their non-inverted model with regards to the data ($1\sigma$ versus $1.5\sigma$). Their conclusion is partly based on WASP-18b's measured temperature of $\sim3200\,$K, and the trend for planets hotter than $2000\,$K to have a stratospheric temperature inversion.}, while \cite{lewis2013} find that HAT-P-2b does have a strong inversion. Both planets have large day to night temperature contrasts \citep{maxted2013,lewis2013}, and both have eclipse brightness temperatures at \three and \four above 2100K \citep{nymeyer2011,lewis2013}. This implies that both planets should have a day-night TiO cold trap that inhibits inversions, though the atmospheric dynamics of HAT-P-2b are complicated by its eccentric (e=0.52) orbit, which causes the stellar insolation to vary considerably.

\cite{parmentier2013} makes the intriguing suggestion that one could test for the presence and efficiency of a day-night TiO cold trap by looking for a latitude dependence in the dayside temperature structure of a planet. Atmospheric gases at higher latitudes could have a shorter nightside crossing-time, lessening the amount of TiO depletion that occurs in the cold trap. If the high-latitudes retain TiO while the equator does not, this would create a latitude-dependent inversion on the dayside of the planet. A latitudinal variation in the dayside temperature could be directly observed by using the phase-mapping technique demonstrated by \cite{majeau2012} and \cite{dewit2012}, though this requires previous knowledge of the longitudinal temperature gradient of the planetary dayside. This argues for obtaining \three and \four orbital phase curve observations of KELT-1b using Spitzer, as those observations would be the only way to directly observe the longitudinal temperature gradient.

\subsection{From a Brown Dwarf Perspective}

As a brown dwarf, KELT-1b is unusual because of the strong stellar insolation it receives, which is far in excess of its own internal luminosity. If it were isolated, we estimate that KELT-1b's surface flux from internal heat should be approximately $10^6$ to $10^7$ erg s$^{-1}$ cm$^{-2}$ \citep{burrows1997}, assuming the discovery paper's age measurement of $\approx$2 Gyr. This heat flux is two to three orders of magnitude less than the incident stellar flux of $7.8\times10^9$ erg s$^{-1}$ cm$^{-2}$. The dayside energy budget of KELT-1b is therefore dominated by the incident stellar radiation. Indeed, our measured brightness temperature of about $3100\,$K corresponds to an mid M-dwarf, even though by its mass, age and surface gravity we would expect KELT-1b to be a mid T-dwarf if it were isolated, at about $700\,$K \citep{burrows2003}. 

The large amount of stellar insolation relative to the internal heat flux of KELT-1b probably means that the atmospheric circulation in KELT-1b is driven by thermal forcing, similar to hot Jupiters. This is in contrast to the atmospheres of cold brown dwarfs, whose circulation is expected to by primarily driven by the breaking of upward-welling gravity waves generated at the radiative-convective boundary. \cite{showman2013} calculate that for internal energy fluxes of $\approx$ $10^8$ erg s$^{-1}$ cm$^{-2}$, which is ten to one hundred times higher than what we expect for KELT-1b based on \cite{burrows1997}, these waves ought to generate atmospheric winds of tens to hundreds of meters per second. By comparison, the thermal forcing in a typical hot Jupiter atmosphere is expected, and observationally implied, to cause wind speeds of several thousand meters per second \citep{knutson2007,snellen2010}. Interestingly, since KELT-1b's rotation is presumably tidally synchronized to its orbital period and thus relatively slow, the atmospheric Rossby number is $Ro=0.25\ (v_{wind}/1\mathrm{km}\ \mathrm{s}^{-1})$(1\rj$/L)$, where $v_{wind}$ is the wind speed and $L$ is the characteristic length scale of atmospheric flows. This is very high compared to the expected Rossby numbers of cold brown dwarfs, which should range from $Ro\approx0.0001$ to $Ro\approx0.01$ \citep{showman2013}. We therefore expect the atmospheric dynamics of KELT-1b to be similar to hot Jupiters.

\subsection{A Combined View?}

Ultimately, we would like to join observations and theories of hot Jupiters and brown dwarfs, by using KELT-1b as one link in that chain. This sort of union is already occurring for brown dwarfs and directly imaged, internal energy dominated, giant planets using the systems discovered around HR 8799 \citep{marois2008} and GJ 504 \citep{kuzuhara2013} as well as the isolated object PSO J318-22 \citep{liu2013}. As of yet, however, there has been no chance to do the same with irradiated Jupiters and brown dwarfs. 

A comparison of irradiated Jupiters and cold brown dwarfs already points to intriguing atmospheric differences between the two populations. Figure 11 shows the equilibrium temperature and $[3.6]-[4.5]$ color of KELT-1b relative to other brown dwarfs and planets. We measure a $[3.6]-[4.5]$ color for KELT-1b of $0.07\pm0.11$. This is consistent with other brown dwarfs at a similar equilibrium temperature, which generally have $[3.6]-[4.5]$ colors near zero for $T_{\rm eq}>1400\,$K. 

The strongly irradiated Jupiters show no such clear behavior with temperature across any of the temperature range. This striking diversity for the planets could be due to a variety of factors, perhaps most strongly influenced by the presence or absence of dayside temperature inversions. Other factors such as non-standard abundance ratios and scatter in Bond albedos and dayside energy redistribution may be important as well. The clear trends in brown dwarf colors with $T_{\rm eff}$ are now well understood in terms of atmospheric chemistry \citep{leggett2010}. At high temperatures H$_2$O and CO absorption bands dominate the infrared spectrum. When temperatures fall below $\sim$$1400\,$K carbon transitions from CO to CH$_4$, which absorbs strongly in the Spitzer \three band. This chemistry change combined with the ever redward shift of the Planck curve to longer wavelengths explains the trend to redder [3.6]-[4.5] Spitzer colors in Figure 11. 

\begin{figure}
\vskip -0.0in 
\epsscale{1.2} 
\plotone{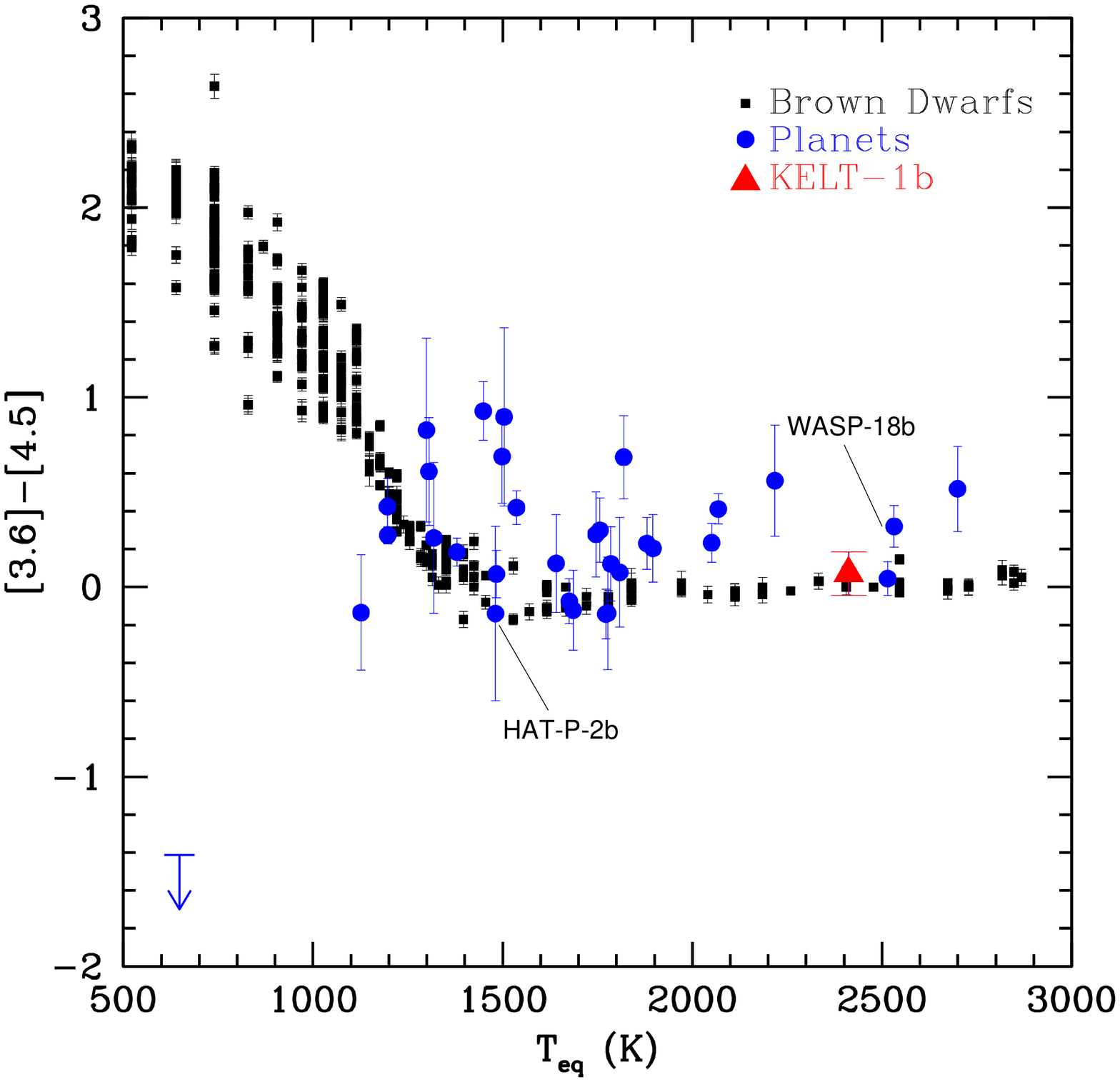}
\vskip -0.0in 
\caption{Spitzer IRAC colors as a function of equilibrium temperature for planets and brown dwarfs. We calculated $T_{\rm eq}$ for the brown dwarfs based on their spectral type and the empirical conversion from spectral type to temperature from \cite{stephens2009}. The brown dwarf colors and spectral types are from \cite{patten2006}, \cite{leggett2010} and \cite{kirkpatrick2011}. The planetary colors are calculated using secondary eclipse depths listed on exoplanets.org and transforming those into fluxes using the $T_{\rm eff}$ of the host star and assuming the host is a blackbody. The planetary $T_{\rm eq}$ values are calculated assuming zero albedo and perfect heat redistribution. The downward arrow is the upper limit for GJ 436b, which has no detected \four eclipse \citep{stevenson2010}..} 
\end{figure}

The fact that KELT-1b so strikingly resembles other brown dwarfs, despite being more like a hot Jupiter in terms of external forcing of the atmosphere by stellar irradiation, suggests that surface gravity is a very important factor in governing the ultimate atmospheric dynamics of these bodies. The further population of this diagram across planet temperature and mass may help identify further differences or similarities between irradiated planets and self-luminous brown dwarfs. 

\section{Summary and Conclusions}

We have measured the secondary eclipse of the highly irradiated transiting brown dwarf KELT-1b in three bands, and found that the object's high surface gravity, and not the high stellar irradiation, dominates KELT-1b's atmosphere. This makes KELT-1b's atmosphere appear more similar to field brown dwarfs at the same effective temperature, rather than to strongly irradiated hot Jupiters. These observations are the first constraints on the atmosphere of a highly irradiated brown dwarf. Specifically, we measure secondary eclipse depths of $0.195\pm0.010$\% at \three and $0.200\pm0.012$\% at 4.5$\,\mu$m. We also find tentative evidence for the secondary eclipse in the $z'$ band with a depth of $0.049\pm0.023$\%. From these measured eclipse depths, we conclude that KELT-1b does not have a high heat redistribution efficiency, and does not show strong evidence for a stratospheric temperature inversion. Importantly, our measurements reveal that KELT-1b has a $[3.6]-[4.5]$ color of $0.07\pm0.11$, identical to that of isolated brown dwarfs of similarly high temperature. In contrast, hot Jupiters generally show redder $[3.6]-[4.5]$ colors of $\sim$0.4, with a very large range from $\sim$0 to $\sim$1. 

KELT-1b gives us the chance to study a high surface gravity atmosphere using all of the tools that have been developed to measure the dynamics in hot Jupiter atmospheres \citep[e.g, HD 189733b by][]{knutson2012}. For the first time we will be able to directly observe large scale atmospheric dynamics and flows in a high surface gravity environment. Already, our secondary eclipse observations demonstrate that there is almost no heat redistribution from the day to the night side of KELT-1b, and suggest that there may be a global hotspot at the substellar point. This implies that the radiative timescale in the atmosphere is extremely short relative to the relevant dynamic timescale in the atmosphere.

By serving as a transitional object between cold brown dwarfs and hot giant planets, more detailed observations of KELT-1b will have a strong ability to illuminate the similarities and differences between these two populations. In the near-term, Spitzer observations of KELT-1b's orbital phase curve would provide the best extension to our current understanding of the object's atmosphere, by directly measuring the day-night temperature contrast and the presence of large scale flows in the atmosphere. Transmission spectroscopy of KELT-1b is currently impossible due to KELT-1b's high surface gravity, though longer-term the James Webb Space Telescope may be able to conduct these observations. Finally, our tentative detection of the eclipse in $z'$ demonstrates the relatively unique opportunities for secondary eclipse observations from the ground that this system affords us -- particularly with modest telescopes.

\acknowledgements
Work by T.G.B. and B.S.G. was partially supported by NSF CAREER grant AST-1056524. K.A.C. was supported by a NASA Kentucky Space Grant Consortium Graduate Fellowship. This research has made use of NASA's Astrophysics Data System.


\begin{thebibliography}{}
\bibitem[Anderson et al.(2011)]{anderson2011} Anderson, D.~R., Collier Cameron, A., Hellier, C., et al.\ 2011, \apjl, 726, L19 

\bibitem[Bakos et al.(2007)]{bakos2007} Bakos, G.~{\'A}., Kov{\'a}cs, G., Torres, G., et al.\ 2007, \apj, 670, 826 

\bibitem[Ballard et al.(2010)]{ballard2010} Ballard, S., Charbonneau, D., Deming, D., et al.\ 2010, \pasp, 122, 1341 

\bibitem[Baskin et al.(2013)]{baskin2013} Baskin, N.~J., Knutson, H.~A., Burrows, A., et al.\ 2013, \apj, 773, 124 

\bibitem[Blecic et al.(2014)]{blecic2014} Blecic, J., Harrington, 
J., Madhusudhan, N., et al.\ 2014, \apj, 781, 116 

\bibitem[Bodenheimer et al.(2013)]{bodenheimer2013} Bodenheimer, P., D'Angelo, G., Lissauer, J.~J., Fortney, J.~J., \& Saumon, D.\ 2013, \apj, 770, 120 

\bibitem[Bouchy et al.(2011a)]{bouchy2011b} Bouchy, F., Bonomo, A.~S., Santerne, A., et al.\ 2011B, \aap, 533, A83

\bibitem[Bouchy et al.(2011b)]{bouchy2011} Bouchy, F., Deleuil, M., Guillot, T., et al.\ 2011A, \aap, 525, A68 

\bibitem[Bowler et al.(2010)]{bowler2010} Bowler, B.~P., Liu, M.~C., Dupuy, T.~J., \& Cushing, M.~C.\ 2010, \apj, 723, 850 

\bibitem[Burgasser(2013)]{burgasser2013} Burgasser, A.~J.\ 2013, Astronomische Nachrichten, 334, 32 

\bibitem[Burrows et al.(1997)]{burrows1997} Burrows, A., Marley, M., Hubbard, W.~B., et al.\ 1997, \apj, 491, 856 

\bibitem[Burrows et al.(2003)]{burrows2003} Burrows, A., Sudarsky, D., \& Lunine, J.~I.\ 2003, \apj, 596, 587   

\bibitem[Cowan \& Agol(2011)]{cowan2011} Cowan, N.~B., \& Agol, E.\ 2011, \apj, 729, 54 

\bibitem[de Wit et al.(2012)]{dewit2012} de Wit, J., Gillon, M., Demory, B.-O., \& Seager, S.\ 2012, \aap, 548, A128 

\bibitem[Deleuil et al.(2008)]{deleuil2008} Deleuil, M., Deeg, H.~J., Alonso, R., et al.\ 2008, \aap, 491, 889 

\bibitem[Demory \& Seager(2011)]{demory2011} Demory, B.-O., \& Seager, S.\ 2011, \apjs, 197, 12 

\bibitem[D{\'{\i}}az et al.(2013)]{diaz2013} D{\'{\i}}az, R.~F., Damiani, C., Deleuil, M., et al.\ 2013, \aap, 551, L9

\bibitem[Eastman et al.(2013)]{eastman2013} Eastman, J., Gaudi, B.~S., \& Agol, E.\ 2013, \pasp, 125, 83 

\bibitem[Eastman et al.(2010)]{eastman2010} Eastman, J., Siverd, R., \& Gaudi, B.~S.\ 2010, \pasp, 122, 935 

\bibitem[Faherty et al.(2013)]{faherty2013} Faherty, J.~K., Cruz, K.~L., Rice, E.~L., \& Riedel, A.\ 2013, \memsai, 84, 955 

\bibitem[Fortney et al.(2008)]{fortney2008} Fortney, J.~J., Lodders, K., Marley, M.~S., \& Freedman, R.~S.\ 2008, \apj, 678, 1419 

\bibitem[Fortney et al.(2006)]{fortney2006} Fortney, J.~J., Saumon, D., Marley, M.~S., Lodders, K., \& Freedman, R.~S.\ 2006, \apj, 642, 495 

\bibitem[Gillon et al.(2007)]{gillon2007} Gillon, M., Demory, B.-O., Barman, T., et al.\ 2007, \aap, 471, L51 

\bibitem[Guillot et al.(1996)]{guillot1996} Guillot, T., Burrows, A., Hubbard, W.~B., Lunine, J.~I., \& Saumon, D.\ 1996, \apjl, 459, L35 

\bibitem[Hellier et al.(2009)]{hellier2009} Hellier, C., Anderson, D.~R., Collier Cameron, A., et al.\ 2009, \nat, 460, 1098 

\bibitem[Hubeny et al.(2003)]{hubeny2003} Hubeny, I., Burrows, A., \& Sudarsky, D.\ 2003, \apj, 594, 1011 

\bibitem[Iro et al.(2005)]{iro2005} Iro, N., B{\'e}zard, B., \& Guillot, T.\ 2005, \aap, 436, 719 

\bibitem[Grether \& Lineweaver(2006)]{grether2006} Grether, D., \& Lineweaver, C.~H.\ 2006, \apj, 640, 1051 

\bibitem[Johns-Krull et al.(2008)]{johnskrull2008} Johns-Krull, C.~M., McCullough, P.~R., Burke, C.~J., et al.\ 2008, \apj, 677, 657 

\bibitem[Johnson et al.(2011)]{johnson2011} Johnson, J.~A., Apps, K., Gazak, J.~Z., et al.\ 2011, \apj, 730, 79  

\bibitem[Kirkpatrick et al.(2011)]{kirkpatrick2011} Kirkpatrick, J.~D., Cushing, M.~C., Gelino, C.~R., et al.\ 2011, \apjs, 197, 19 

\bibitem[Knutson et al.(2007)]{knutson2007} Knutson, H.~A., Charbonneau, D., Allen, L.~E., et al.\ 2007, \nat, 447, 183 

\bibitem[Knutson et al.(2008)]{knutson2008} Knutson, H.~A., Charbonneau, D., Allen, L.~E., Burrows, A., \& Megeath, S.~T.\ 2008, \apj, 673, 526

\bibitem[Knutson et al.(2012)]{knutson2012} Knutson, H.~A., Lewis, N., Fortney, J.~J., et al.\ 2012, \apj, 754, 22  

\bibitem[Kuzuhara et al.(2013)]{kuzuhara2013} Kuzuhara, M., Tamura, M., Kudo, T., et al.\ 2013, \apj, 774, 11 

\bibitem[Lagrange et al.(2010)]{lagrange2010} Lagrange, A.-M., Bonnefoy, M., Chauvin, G., et al.\ 2010, Science, 329, 57 

\bibitem[Leggett et al.(2010)]{leggett2010} Leggett, S.~K., Burningham, B., Saumon, D., et al.\ 2010, \apj, 710, 1627 

\bibitem[Lewis et al.(2013)]{lewis2013} Lewis, N.~K., Knutson, H.~A., Showman, A.~P., et al.\ 2013, \apj, 766, 95   

\bibitem[Liu et al.(2013)]{liu2013} Liu, M.~C., Magnier, E.~A., Deacon, N.~R., et al.\ 2013, \apjl, 777, L20

\bibitem[Lucy \& Sweeney(1971)]{lucy1971} Lucy, L.~B., \& Sweeney, M.~A.\ 1971, \aj, 76, 544 

\bibitem[Luhman(2012)]{luhman2012} Luhman, K.~L.\ 2012, \araa, 50, 65 

\bibitem[Majeau et al.(2012)]{majeau2012} Majeau, C., Agol, E., \& Cowan, N.~B.\ 2012, \apjl, 747, L20 

\bibitem[Mandel \& Agol(2002)]{mandel2002} Mandel, K., \& Agol, E.\ 2002, \apjl, 580, L171

\bibitem[Marcy \& Butler(2000)]{marcy2000} Marcy, G.~W., \& Butler, R.~P.\ 2000, \pasp, 112, 137 

\bibitem[Marley et al.(2010)]{marley2010} Marley, M.~S., Saumon, D., \& Goldblatt, C.\ 2010, \apjl, 723, L117 

\bibitem[Marois et al.(2008)]{marois2008} Marois, C., Macintosh, B., Barman, T., et al.\ 2008, Science, 322, 1348 

\bibitem[Maxted et al.(2013)]{maxted2013} Maxted, P.~F.~L., Anderson, D.~R., Doyle, A.~P., et al.\ 2013, \mnras, 428, 2645 

\bibitem[Mighell(2005)]{mighell2005} Mighell, K.~J.\ 2005, \mnras, 361, 861 

\bibitem[Miller \& Fortney(2011)]{miller2011} Miller, N., \& Fortney, J.~J.\ 2011, \apjl, 736, L29 

\bibitem[Molli{\`e}re \& Mordasini(2012)]{molliere2012} Molli{\`e}re, P., \& Mordasini, C.\ 2012, A\&A, 547, A105  

\bibitem[Moutou et al.(2013)]{moutou2013} Moutou, C., Bonomo, A.~S., Bruno, G., et al.\ 2013, \aap, 558, L6  

\bibitem[Moutou et al.(2004)]{moutou2004} Moutou, C., Pont, F., Bouchy, F., \& Mayor, M.\ 2004, \aap, 424, L31

\bibitem[Nymeyer et al.(2011)]{nymeyer2011} Nymeyer, S., Harrington, J., Hardy, R.~A., et al.\ 2011, \apj, 742, 35 

\bibitem[Parmentier et al.(2013)]{parmentier2013} Parmentier, V., Showman, A.~P., \& Lian, Y.\ 2013, \aap, 558, A91 

\bibitem[Patten et al.(2006)]{patten2006} Patten, B.~M., Stauffer, J.~R., Burrows, A., et al.\ 2006, \apj, 651, 502 

\bibitem[Perez-Becker \& Showman(2013)]{perez2013} Perez-Becker, D., \& Showman, A.~P.\ 2013, \apj, 776, 134 

\bibitem[Sahlmann et al.(2011)]{sahlmann2011} Sahlmann, J., S{\'e}gransan, D., Queloz, D., et al.\ 2011, \aap, 525, A95

\bibitem[Seager(2010)]{seager2010} Seager, S.\ 2010, Exoplanet Atmospheres: Physical Processes (Princeton, NJ: Princeton Univ. Press)

\bibitem[Showman et al.(2008)]{showman2008} Showman, A.~P., Cooper, C.~S., Fortney, J.~J., \& Marley, M.~S.\ 2008, \apj, 682, 559 

\bibitem[Showman et al.(2009)]{showman2009} Showman, A.~P., Fortney, J.~J., Lian, Y., et al.\ 2009, \apj, 699, 564 

\bibitem[Showman \& Kaspi(2013)]{showman2013} Showman, A.~P., \& Kaspi, Y.\ 2013, \apj, 776, 85 

\bibitem[Siverd et al.(2012)]{siverd2012} Siverd, R.~J., Beatty, T.~G., Pepper, J., et al.\ 2012, \apj, 761, 123 

\bibitem[Snellen et al.(2010)]{snellen2010} Snellen, I.~A.~G., de Kok, R.~J., de Mooij, E.~J.~W., \& Albrecht, S.\ 2010, \nat, 465, 1049  

\bibitem[Spiegel et al.(2011)]{spiegel2011} Spiegel, D.~S., Burrows, A., \& Milsom, J.~A.\ 2011, \apj, 727, 57 

\bibitem[Stassun et al.(2006)]{stassun2006} Stassun, K.~G., Mathieu, R.~D., \& Valenti, J.~A.\ 2006, \nat, 440, 311 

\bibitem[Stassun et al.(2007)]{stassun2007} Stassun, K.~G., Mathieu, R.~D., \& Valenti, J.~A.\ 2007, \apj, 664, 1154 

\bibitem[Stephens et al.(2009)]{stephens2009} Stephens, D.~C., Leggett, S.~K., Cushing, M.~C., et al.\ 2009, \apj, 702, 154 

\bibitem[Stevenson et al.(2010)]{stevenson2010} Stevenson, K.~B., Harrington, J., Nymeyer, S., et al.\ 2010, \nat, 464, 1161 
\end{thebibliography}
\end{document}